\pdfoutput=1
\documentclass[a4paper]{article}
\usepackage{graphicx}
\usepackage{fullpage}
\usepackage{amsmath}
\newcommand{\ve}{\mathbf}
\newcommand{\T}{\mathsf{T}}

\begin{document}

\title{Explicitly correlated Gaussians with tensor pre-factors:
analytic matrix elements}

\author{
        D.V.~Fedorov (email: fedorov@phys.au.dk),\\
        A.F.~Teilmann, M.C.~Østerlund, T.L.~Norrbohm \\
        {\it Aarhus University, Aarhus, Denmark}
}

\date{}
\maketitle

\begin{abstract}
We consider a specific form of explicitly correlated Gaussians---with
tensor pre-factors---which appear naturally when dealing with certain
few-body systems in nuclear and particle physics.  We derive analytic
matrix elements with these Gaussians---overlap, kinetic energy,
and Coulomb potential---to be used in variational calculations of
those systems.  We also perform a quick test of the derived formulae by
applying them to p- and d-waves of the hydrogen atom.  \end{abstract}

\section{Introduction}
Explicitly correlated Gaussians in various forms is a popular
basis for variational calculations of quantum-mechanical
few-body systems~\cite{book,hyama,review}. One important
advantage of Gaussians is that their matrix elements are often
analytic~\cite{hyama,complex,L1,angular,shifted} which greatly facilitates
the calculations.

The most general Gaussians are shifted correlated Gaussians.  However,
they have a disadvantage: they lack a definite angular momentum.
That might be problematic in applications where angular momentum is
an important quantum number. In the literature one can find analytic
formulae with Gaussians that acquire a definite angular momentum via
pre-factors built of spherical harmonics~\cite{angular}. However there
exist applications in nuclear and particle physics, in particular with
pions~\cite{aksel,gammapi,roper}, where the variational basis-functions
are multiplied by operators in the form of vectors and tensors\footnote{
For example, the operators $\vec\sigma\vec r$ and $\vec\sigma\vec p$,
where $\vec\sigma$ are Pauli matrices, $\vec r$ and $\vec p$ are the
coordinate and momentum of the pion.}.  These applications favour
Gaussians with vector and tensor pre-factors rather than spherical
harmonics. In this contribution we derive the analytic matrix elements
with such Gaussians. We start with the matrix elements of the shifted
correlated Gaussians, perform their Taylor expansion with respect to
shift-vectors, and then collect the terms of specific orders, which
ultimately gives the sought matrix elements.

Since some of the resulting formulae are rather long, particularly for
the tensor pre-factor, we perform a quick test of the derived formulae
by applying them to the lowest p- and d-wave states of the hydrogen atom.

\section{Shifted Gaussians}
Since we are going to use---as the starting point
of our derivations---the analytic matrix elements with shifted
correlated Gaussians
we reproduce here the relevant analytic formulae
from~\cite{shifted}.

For an $n$-body system of particles the shifted correlated Gaussian
with correlation
matrix $A$ and shift $\ve a$ is defined in coordinate space $\ve r$ as
	\begin{equation}
\langle \ve r|A,\ve a\rangle \doteq
		e^{-\ve r^\T A\ve r+\ve a^\T\ve r} \;,
	\end{equation}
where ${}^\T$ denotes transposition,
$\ve r$ is the column
$\{\vec r_1,\dots,\vec r_n\}$
of the coordinates of the bodies,
the shift $\ve a$ is a column of shift vectors
$\{\vec a_1,\dots,\vec a_n\}$,
and where
	\begin{eqnarray}
\ve r^\T A\ve r \doteq \sum_{i,j=1}^{n}\vec r_i\cdot A_{ij}\vec r_j \,,\\
\ve a^\T\ve r \doteq \sum_{i=1}^n \vec a_i\cdot\vec r_i \,,
	\end{eqnarray}
where the central dot denotes the scalar product of two vectors.

The overlap of two shifted Gaussians is given as
	\begin{equation}\label{genover}
\langle B,\ve b|A,\ve a\rangle =
e^{\frac14(\ve a+\ve b)^\T R(\ve a+\ve b)}
 \left(
	\frac{\pi^{n}}{\mathrm{det}(A+B)}
	\right)^{3/2}
\doteq M \doteq e^{\frac14(\ve a+\ve b)^\T R(\ve a+\ve b)}M_0 \,,
	\end{equation}
where $R=(A+B)^{-1}$.

The matrix element of the kinetic energy operator is given as
	\begin{eqnarray}\label{genkin}
&
\langle B,\ve b|\left(-\frac{\partial}{\partial\ve r} K
\frac{\partial}{\partial\ve r^\T}\right)|A,\ve a\rangle =
6\,\mathrm{Tr}(BKAR)M
+\ve b^\T K\ve a M
& \nonumber\\
&
+(\ve a+\ve b)^\T RBKAR (\ve a+\ve b) M
-(\ve a+\ve b)^\T RBK \ve a M
-\ve b^\T KAR (\ve a+\ve b) M \,,
&
	\end{eqnarray}
where $K$ is the (reduced) mass-matrix of the system of particles in
the given set of coordinates.

The matrix element of the Coulomb potential is given as
	\begin{equation}\label{gencoul}
\langle B,\ve b| \frac{1}{|w^\T\ve r|} |A,\ve a\rangle =
\frac{\mathrm{erf}(\sqrt{\beta}q)}{q} M \,,
	\end{equation}
where $w$ is a given column of numbers
$\{w_1,\dots,w_n\}$, and where
$\beta\doteq(w^\T Rw)^{-1}$, $\vec q\doteq\frac12 w^\T R(\ve a+\ve b)$.

\section{Rank-0 Gaussians}
The matrix elements with rank-0 (s-wave) Gaussians are well
known. Notwithstanding, we reproduce here the relevant formulae for
completeness.
A rank-0 Gaussian is the zero-shift limit of a shifted Gaussian,
	\begin{equation}
\langle \ve r|A \rangle \doteq \lim_{\ve a\to 0}\langle\ve r|A,\ve
a\rangle = e^{-\ve r^\T A\ve r} \;.
	\end{equation}

The overlap of two rank-0 Gaussians is the zero-shift limit of the
shifted overlap~(\ref{genover}),
	\begin{equation}
\langle B|A\rangle = \lim_{\ve a,\ve b\to 0}M \doteq M_0 = \left(
	\frac{\pi^{n}}{\mathrm{det}(A+B)}
	\right)^{3/2} \;.
	\end{equation}

Kinetic energy matrix element with rank-0 Gaussians is again given as
the zero-shift limit of the corresponding shifted
matrix element~(\ref{genkin}),
	\begin{equation}
\langle B|
\left(
-\frac{\partial}{\partial\ve r}K \frac{\partial}{\partial\ve r^\T}
\right)
\left|A\right\rangle =
\lim_{\ve a,\ve b\to 0}
\langle B,\ve b|
\left(
-\frac{\partial}{\partial\ve r} K \frac{\partial}{\partial\ve r^\T}
\right)
|A,\ve a\rangle =
6\mathrm{Tr}(BKAR)
M_0
\;.
	\end{equation}

The Coulomb potential matrix element is, analogously, the zero-shift limit
of the shifted matrix element~(\ref{gencoul}),
	\begin{equation}
\langle B|\frac{1}{|w^\T\ve r|}|A\rangle =
\lim_{q\rightarrow 0} \frac{\mathrm{erf}(\sqrt{\beta}q)}{q}
\lim_{\ve a,\ve b\to 0}M =
2\sqrt{\frac{\beta}\pi} M_0
\;.
	\end{equation}

\section{Rank-1 pre-factor Gaussians}
The rank-1 pre-factor Gaussians are constructed by pre-factoring rank-0
Gaussians with the form $(\ve a^\T\ve r)$ (which
represents a pure p-wave). We shall use the following notation,
	\begin{equation}
\langle \ve r|(\ve a)A\rangle
\doteq (\ve a^\T\ve r)e^{-\ve r^\T A\ve r} \;,
	\end{equation}
where $\ve a$ is a column of polarization vectors $\{\vec a_1,\dots,\vec
a_n\}$, and $|(\ve a)A\rangle$ is our rank-1 Gaussian.

The rank-1 Gaussian can be obtained by collecting the linear term in the
Taylor expansion of the shifted Gaussian in the shift-variable,
	\begin{equation}
\langle \ve r|A,\ve a\rangle
\doteq e^{-\ve r^\T A\ve r+\ve a^\T\ve r}
= e^{-\ve r^\T A\ve r}+(\ve a^\T\ve r)e^{-\ve r^\T A\ve r}
+\frac12(\ve a^\T\ve r)^2e^{-\ve r^\T A\ve r}+\dots \;.
	\end{equation}
We shall denote this by the following notation,
	\begin{equation}
\langle \ve r|A,\ve a\rangle 
\underset{O(\ve a)}{\longrightarrow}
(\ve a^\T\ve r)e^{-\ve r^\T A\ve r}\doteq |(\ve a)A\rangle \;.
	\end{equation}

The overlap of rank-1 Gaussians can be obtained by collecting the terms on
the order $O(\ve a\ve b)$ from the Taylor expansions of the shifted
overlap
(and employing the symmetry of the matrix $R$),
	\begin{eqnarray}
\left\langle B,\ve b\right| \left. A,\ve a\right\rangle
=
\left\langle e^{-\ve r^\T B\ve r+\ve b^\T \ve r} \right|
\left. e^{-\ve r^\T A\ve r+\ve a^\T \ve r}\right\rangle
\underset{O(\ve a\ve b)}{\longrightarrow}
\left\langle (\ve b^\T \ve r)e^{-\ve r^\T B\ve r} \right|
\left. (\ve a^\T \ve r)e^{-\ve r^\T A\ve r}\right\rangle
=
\left\langle (\ve b)  B \right| \left.  (\ve a) A \right\rangle
	\;. \end{eqnarray}
The $O(\ve a\ve b)$ term of the expansion is given as
	\begin{equation}
\left\langle B,\ve b\right| \left. A,\ve a\right\rangle \doteq M=
e^{\frac14(\ve a+\ve b)^\T R(\ve a+\ve b)}M_0
\underset{O(\ve a\ve b)}{\longrightarrow}
\frac12(\ve b^\T R\ve a)M_0 \;,
	\end{equation}
which leads to
	\begin{equation}
\big\langle (\ve b) B\big|(\ve a) A\big\rangle 
\doteq M_1
= \frac{1}{2}(\ve b^\T R \ve a) M_0
	\;. \end{equation}

Similarly, the  kinetic energy matrix element with rank-1 Gaussians can be
obtained by collecting the terms on the order $O(\ve a\ve b)$ from
the Taylor expansion of the corresponding shifted Gaussian matrix element,
	\begin{equation}
\langle B,\ve b|
\left(
-\frac{\partial}{\partial\ve r}K\frac{\partial}{\partial\ve r^\T}
\right)
|A,\ve a\rangle
\underset{O(\ve a\ve b)}{\longrightarrow}
\langle (\ve b)B|
\left(
-\frac{\partial}{\partial\ve r}K\frac{\partial}{\partial\ve r^\T}
\right)
|(\ve a)A\rangle \;.
	\end{equation}
The Taylor expansion of the shifted kinetic energy matrix
element~(\ref{genkin}) gives, term by term,
	\begin{equation}
6\mathrm{Tr}(BKAR)M
\underset{O(\ve a\ve b)}{\longrightarrow}
6\mathrm{Tr}(BKAR)M_1 \;,
	\end{equation}
	\begin{equation}
\ve b^\T K\ve a M
\underset{O(\ve a\ve b)}{\longrightarrow}
\ve b^\T K\ve a M_0 \;,
	\end{equation}
	\begin{equation}
(\ve a+\ve b)^\T RBKAR(\ve a+\ve b) M
\underset{O(\ve a\ve b)}{\longrightarrow}
  \ve a^\T RBKAR \ve b M_0
+ \ve b^\T RBKAR \ve a M_0 \;,
	\end{equation}
	\begin{equation}
-(\ve a+\ve b)^\T RBK \ve a M
\underset{O(\ve a\ve b)}{\longrightarrow}
-\ve b^\T RBK \ve a M_0 \;,
	\end{equation}
	\begin{equation}
-(\ve a+\ve b)^\T RAK \ve b M
\underset{O(\ve a\ve b)}{\longrightarrow}
-\ve a^\T RAK \ve b M_0 \;.
	\end{equation}
Summing up,
	\begin{eqnarray}
&&\langle (\ve b)B|
\left(-\frac{\partial}{\partial\ve r}K
\frac{\partial}{\partial\ve r^\T}\right)
\left|(\ve a)A\right\rangle =
6\mathrm{Tr}(BKAR)M_1 \nonumber\\
&&+\ve b^\T K\ve a M_0
+\ve a^\T RBKAR \ve b M_0
+ \ve b^\T RBKAR \ve a M_0
-\ve b^\T RBK \ve a M_0
-\ve a^\T RAK \ve b M_0 \;.
	\end{eqnarray}

The Coulomb matrix element is again obtained by collecting the terms
$O(\ve a\ve b)$ from the Taylor expansion of the shifted Coulomb matrix
element~(\ref{gencoul}),
	\begin{equation}
\left\langle B,\ve b\right| \frac{1}{|w^\T \ve r|} \left|A,\ve a\right\rangle
\underset{O(\ve a\ve b)}{\longrightarrow}
\left\langle (\ve b)B\right|
\frac{1}{|w^\T \ve r|}
\left|(\ve a)A\right\rangle \;.
	\end{equation}
We first expand the error-function,\footnote{
	\begin{equation}
\frac{\mathrm{erf}(z)}{z}= \frac{2}{\sqrt\pi}
\left(
1-\frac{z^2}{3}+\frac{z^4}{10}-\frac{z^6}{42}+\frac{z^8}{216}-\cdots
\right)
	\end{equation}
}
	\begin{eqnarray}
\left\langle B,\ve b\right|
\frac{1}{|w^\T \ve r|}
\left|A,\ve a\right\rangle
=\frac{\mathrm{erf}(\sqrt\beta q)}{\sqrt{\beta}q}\sqrt{\beta}M
=
2\sqrt{\frac{\beta}{\pi}}\left(1-\frac{\beta q^2}{3}+\dots\right)
e^{\frac14(\ve a+\ve b)^\T R(\ve a+\ve b)}M_0 \;,
	\end{eqnarray}
where
	\begin{equation}
q^2 = \frac14(\ve a+\ve b)^\T Rww^\T R(\ve a+\ve b) \;.
	\end{equation}
Now, performing full expansion and collecting the terms
$O(\ve a\ve b)$, gives
	\begin{equation}
\left\langle B,\ve b\right|
\frac{1}{|w^\T \ve r|}
\left|A,\ve a\right\rangle
\underset{O(\ve a\ve b)}{\longrightarrow}
2\sqrt{\frac{\beta}{\pi}} M_1
-\sqrt{\frac{\beta}{\pi}}\frac{\beta}{3}\ve b^\T Rww^\T R \ve a M_0
\doteq
\left\langle (\ve b)B\right|
\frac{1}{|w^\T \ve r|}
\left|(\ve a)A\right\rangle \;.
	\end{equation}

\section{Rank-2 pre-factor Gaussians}

The rank-2 pre-factor Gaussians are constructed as
	\begin{equation}
\langle \ve r|(\ve a\ve b)A\rangle
\doteq (\ve a^\T\ve r)(\ve b^\T\ve r)e^{-\ve r^\T A\ve r} \;,
	\end{equation}
where $\ve a$ and $\ve b$ are the polarization vectors. The rank-2
pre-factor generally contains both s-waves and d-waves, however the
condition $\ve a^\T\ve b=0$ eliminates the s-wave contribution.

The rank-2 Gaussian is the $O(\ve a\ve b)$ term in the Taylor expansion
of the shifted Gaussian,
	\begin{equation}
\langle \ve r|A,\ve a+\ve b\rangle
=e^{\ve r^\T A\ve r+(\ve a+\ve b)^\T\ve r}
\underset{O(\ve a\ve b)}{\longrightarrow}
(\ve a^\T\ve r)(\ve b^\T\ve r)e^{-\ve r^\T A\ve r}
\doteq \langle \ve r|(\ve a\ve b)A\rangle \;,
	\end{equation}

The overlap of rank-2 Gaussians is given by the terms $O(\ve a\ve b\ve
c\ve d)$ from
the Taylor expansion of the shifted overlap,
	\begin{eqnarray}
\left\langle B,\ve c+\ve d\right|\left. A,\ve a+\ve b\right\rangle
&=&
\left\langle e^{(\ve c+\ve d)^\T\ve r+\ve r^\T B\ve r}\right|
\left.e^{(\ve a+\ve b)^\T\ve r+\ve r^\T A\ve r}\right\rangle
\nonumber\\
&\underset{O(\ve a\ve b\ve c\ve d)} \longrightarrow &
\left\langle
(\ve c^\T\ve r)(\ve d^\T\ve r)e^{\ve r^\T B\ve r}
\right|\left.
(\ve a^\T\ve r)(\ve b^\T\ve r)e^{\ve r^\T A\ve r}
\right\rangle
=
\left\langle (\ve c\ve d)B \right| \left. (\ve a\ve b)B \right\rangle
\;.
	\end{eqnarray}
Performing the Taylor expansion,
	\begin{eqnarray}
\left\langle B,\ve c+\ve d\right|\left. A,\ve a+\ve b\right\rangle
&=&
e^{\frac14(\ve a+\ve b+\ve c+\ve d)^\T R(\ve a+\ve b+\ve c+\ve d)}M_0
\nonumber \\
&\underset{O(\ve a\ve b\ve c\ve d)} \longrightarrow &
\frac14\Big[
 (\ve a^\T R\ve b)(\ve c^\T R\ve d)
+(\ve a^\T R\ve c)(\ve b^\T R\ve d)
+(\ve a^\T R\ve d)(\ve b^\T R\ve c)
\Big]M_0 \;,
	\end{eqnarray}
gives
	\begin{equation}
\left\langle (\ve c\ve d)B \right| \left. (\ve a\ve b)A \right\rangle
\doteq M_2=
\frac14\Big[
 (\ve a^\T R\ve b)(\ve c^\T R\ve d)
+(\ve a^\T R\ve c)(\ve b^\T R\ve d)
+(\ve a^\T R\ve d)(\ve b^\T R\ve c)
\Big]M_0 \;.
	\end{equation}

\subsection{Kinetic energy}
The rank-2 kinetic energy matrix element is given by the sum of
the $O(\ve a\ve b\ve c\ve d)$ terms in the Taylor expansion of the
shifted Gaussian kinetic energy~(\ref{genkin}). Performing the Taylor
expansion of~(\ref{genkin}) gives the following, term by term,
\begin{enumerate}
\item the $6~\mathrm{Tr}(BKAR)M$ term,
	\begin{equation}
6~\mathrm{Tr}(BKAR)M\underset{O(\ve a\ve b\ve c\ve d)} \longrightarrow
6~\mathrm{Tr}(BKAR)M_2
	\;.\end{equation}
\item the $\ve b^\T K\ve aM$ term,
	\begin{eqnarray}
(\ve c+\ve d)^\T K(\ve a+\ve b)
e^{\frac14(\ve a+\ve b+\ve c+\ve d)^\T R(\ve a+\ve b+\ve c+\ve d)}M_0
\underset{O(\ve a\ve b\ve c\ve d)} \longrightarrow \nonumber\\
\left[
 (\ve a^\T K\ve c)(\ve b^\T R\ve d)
+(\ve a^\T K\ve d)(\ve b^\T R\ve c)
+(\ve b^\T K\ve c)(\ve a^\T R\ve d)
+(\ve b^\T K\ve d)(\ve a^\T R\ve c)
\right] \frac12 M_0
	\;.\end{eqnarray}
\item the $(\ve a+\ve b)^\T RBKAR (\ve a+\ve b)M$ term,
	\begin{eqnarray}
(\ve a+\ve b+\ve c+\ve d)^\T RBKAR(\ve a+\ve b+\ve c+\ve d)
e^{\frac14(\ve a+\ve b+\ve c+\ve d)^\T R(\ve a+\ve b+\ve c+\ve d)}M_0
\underset{O(\ve a\ve b\ve c\ve d)} \longrightarrow \nonumber\\
\left[
 (\ve a^\T RBKAR \ve b)(\ve c R\ve d)
+(\ve a^\T RBKAR \ve c)(\ve b R\ve d)
+(\ve a^\T RBKAR \ve d)(\ve b R\ve c)
\right] \frac12 M_0 \nonumber \\
+\left[
 (\ve b^\T RBKAR \ve a)(\ve c R\ve d)
+(\ve b^\T RBKAR \ve c)(\ve a R\ve d)
+(\ve b^\T RBKAR \ve d)(\ve a R\ve c)
\right] \frac12 M_0 \nonumber \\
+\left[
 (\ve c^\T RBKAR \ve a)(\ve b R\ve d)
+(\ve c^\T RBKAR \ve b)(\ve a R\ve d)
+(\ve c^\T RBKAR \ve d)(\ve a R\ve b)
\right] \frac12 M_0 \nonumber \\
+\left[
 (\ve d^\T RBKAR \ve a)(\ve b R\ve c)
+(\ve d^\T RBKAR \ve b)(\ve a R\ve c)
+(\ve d^\T RBKAR \ve c)(\ve a R\ve b)
\right] \frac12 M_0
	\;.\end{eqnarray}
\item the $-(\ve a+\ve b)^\T RBK \ve aM$ term,
	\begin{eqnarray}
-(\ve a+\ve b+\ve c+\ve d)^\T RBK (\ve a+\ve b)
e^{\frac14(\ve a+\ve b+\ve c+\ve d)^\T R(\ve a+\ve b+\ve c+\ve d)}M_0
\underset{O(\ve a\ve b\ve c\ve d)} \longrightarrow \nonumber\\
-\left[
 (\ve a^\T RBK \ve b)(\ve c^\T R\ve d)
+(\ve b^\T RBK \ve a)(\ve c^\T R\ve d)
\right] \frac12 M_0 \nonumber\\
-\left[
 (\ve c^\T RBK \ve a)(\ve b^\T R\ve d)
+(\ve c^\T RBK \ve b)(\ve a^\T R\ve d)
\right] \frac12 M_0 \nonumber\\
-\left[
 (\ve d^\T RBK \ve a)(\ve b^\T R\ve c)
+(\ve d^\T RBK \ve b)(\ve c^\T R\ve a)
\right] \frac12 M_0
	\;.\end{eqnarray}
\item the $-\ve b^\T KAR (\ve a+\ve b)M$ term,
	\begin{eqnarray}
-(\ve c+\ve d)^\T KAR(\ve a+\ve b+\ve c+\ve d)
e^{\frac14(\ve a+\ve b+\ve c+\ve d)^\T R(\ve a+\ve b+\ve c+\ve d)}M_0
\underset{O(\ve a\ve b\ve c\ve d)} \longrightarrow \nonumber\\
-\left[
 (\ve c^\T KAR \ve a)(\ve b^\T R\ve d)
+(\ve c^\T KAR \ve b)(\ve a^\T R\ve d)
+(\ve c^\T KAR \ve d)(\ve a^\T R\ve b)
\right] \frac12 M_0 \nonumber\\
-\left[
 (\ve d^\T KAR \ve a)(\ve b^\T R\ve c)
+(\ve d^\T KAR \ve b)(\ve a^\T R\ve c)
+(\ve d^\T KAR \ve c)(\ve a^\T R\ve b)
\right] \frac12 M_0
	\;.\end{eqnarray}
\end{enumerate}

\subsection{Coulomb}
The rank-2 Coulomb matrix element is given by the sum of the terms
$O(\ve a\ve b\ve c\ve d)$ in the Taylor
expansion of the shifted Gaussian Coulomb matrix element~(\ref{gencoul}),
	\begin{eqnarray}
&&\left\langle B,\ve c+\ve d\right|
\frac{1}{|w^\T \ve r|}
\left|\ve A,\ve a+\ve b\right\rangle
=\frac{\mathrm{erf}(\sqrt\beta q)}{\sqrt{\beta}q}\sqrt{\beta}M
\nonumber\\
&&=
2\sqrt{\frac{\beta}{\pi}}
\left(1-\frac{\beta q^2}{3}+\frac{\beta^2q^4}{10}+\dots\right)
e^{\frac14(\ve a+\ve b+\ve c+\ve d)^\T R(\ve a+\ve b+\ve c+\ve d)}M_0
	\;,\end{eqnarray}
where
	\begin{equation}
q^2 =
\frac14(\ve a+\ve b+\ve c+\ve d)^\T Rww^\T R
(\ve a+\ve b+\ve c+\ve d) \,.
	\end{equation}
Performing the expansion and collecting the $O(\ve a\ve b\ve c\ve d)$
terms gives
	\begin{eqnarray}
&
\left\langle (\ve c\ve d)B\right|
\frac{1}{|w^\T \ve r|}
\left|(\ve a\ve b)A\right\rangle
=
2\sqrt{\frac{\beta}{\pi}}M_2
&
\nonumber\\
&
-2\sqrt{\frac{\beta}{\pi}}\frac{\beta}{3}\frac14 M_0
\left[\begin{array}{c}
~~(\ve a^\T Rww^\T R \ve b)(\ve c^\T R\ve d) \\
+(\ve a^\T Rww^\T R \ve c)(\ve b^\T R\ve d) \\
+(\ve a^\T Rww^\T R \ve d)(\ve b^\T R\ve c) \\
+(\ve b^\T Rww^\T R \ve c)(\ve a^\T R\ve d) \\
+(\ve b^\T Rww^\T R \ve d)(\ve a^\T R\ve c) \\
+(\ve c^\T Rww^\T R \ve d)(\ve a^\T R\ve b)
\end{array}\right]
+2\sqrt{\frac{\beta}{\pi}}\frac{\beta^2}{10}\frac12 M_0
\left[\begin{array}{c}
~~(\ve a^\T Rww^\T R \ve b)(\ve c^\T Rww^\T R \ve d) \\
 +(\ve a^\T Rww^\T R \ve c)(\ve b^\T Rww^\T R \ve d) \\
 +(\ve a^\T Rww^\T R \ve d)(\ve b^\T Rww^\T R \ve c) \\
\end{array}\right] .
&
	 \end{eqnarray}

\section{Symmetrization}
The wave-function of a few-body system must be properly (anti)symmetrized
under the exchange of the coordinates of identical particles. The
total (anti)symmetrization
can be done by a proper combination of the (elementary) permutation
operators $\hat P$
that exchange the coordinates of the particles,
$\vec r_{i}\to\vec r_{i'}$, such that
	\begin{equation}
\ve r\to P\ve r \;,
	\end{equation}
where the matrix $P$ has the matrix elements
	\begin{equation}
(P)_{ij} = \delta_{ji'} \;.
	\end{equation}
Under this operation the terms $\ve r^\T A\ve r$ and $\ve
s^\T \ve r$ transform as
	\begin{eqnarray}
\ve r^\T A\ve r &\to& \ve r^\T (P^\T A P) \ve r \;, \\
\ve s^\T \ve r &\to& (P^\T\ve s) \ve r \;.
	\end{eqnarray}
Consequently the action of the permutation operator on the
shifted/pre-factor Gaussians is given as
	\begin{eqnarray}
\hat P\left|A,\ve s\right\rangle &=& \left|(P^\T A P), (P^\T\ve
a)\right\rangle \;, \\
\hat P\left|(\ve a)A\right\rangle &=& \left|(P^\T\ve a)(P^\T A
P)\right\rangle \;, \\
\hat P\left|(\ve a\ve b)A\right\rangle &=& \left|(P^\T\ve a~P^\T\ve b)(P^\T
A P)\right\rangle \;.
	\end{eqnarray}
That is, under the permutation operation the Gaussians transform into
the Gaussians of the same functional form---with the same analytic
formulae for the matrix elements---only the correlation matrix and the
shift/pre-factor vectors must be modified appropriately.

\section{Quick test}
As a quick test of the derived formulae
we calculate the lowest energies for s-, p-, and d-waves of the hydrogen
atom by solving the corresponding Schrodinger equation,
	\begin{equation}
\hat H\Psi_l(\vec r)=E_l\Psi_l(\vec r) \;,
	\end{equation}
where the Hamiltonian operator---kinetic energy plus Coulomb
potential---is given (in Hartree units) as
	\begin{equation}
\hat H = -\frac12\frac{\partial^2}{\partial\vec r^2} -\frac{1}{r} \;.
	\end{equation}
The wave-function $\Psi_l(\vec r)$ with the given angular momentum~$l$
is represented as an expansion in terms of a size-$n$ set of Gaussians
with pre-factors corresponding to the given angular momentum,
	\begin{equation}
\Psi_l=\sum_{i=1}^{n}c_i^{(l)} |G_i^{(l)}\rangle \;,
	\end{equation}
where $c_i^{(l)}$ are the expansion coefficients, and where
$|G^{(l)}\rangle$ is a Gaussian with a tensor pre-factor of rank-$l$.

The rank-0 (s-wave) Gaussians are given as
	\begin{equation}
\langle\vec r|G_i^{(0)}\rangle
=\langle\vec r|A_i\rangle
=e^{-\alpha_i r^2} \;,
	\end{equation}
where $\alpha_i$ is the variational parameter.
The rank-1 (p-wave) Gaussians are taken as
	\begin{equation}
\langle\vec r|G_i^{(1)}\rangle
=\langle\vec r|(\vec a)A_i\rangle
=(\vec a\cdot\vec r)e^{-\alpha_i r^2} \;,
	\end{equation}
where the polarization vector $\vec a$ is chosen to be $\vec a=\{0,0,1\}$
such that
	\begin{equation}
\vec a\vec r=z\propto rY_{10}(\vec r) \;.
	\end{equation}
The rank-2 Gaussians are taken in the form
	\begin{equation}
\langle\vec r|G_i^{(2)}\rangle
=\langle\vec r|(\vec a\vec b)A_i\rangle
=(\vec a\cdot\vec r)(\vec b\cdot\vec r)e^{-\alpha_i r^2} \;,
	\end{equation}
where the polarization vectors are chosen as
$\vec a=\{1,0,0\}$,
$\vec b=\{0,1,0\}$
such that $\vec a\cdot\vec b=0$ and the Gaussian represents a pure
d-wave,
	\begin{equation}
(\vec a\vec r)(\vec b\vec r)=xy
\propto r^2\Big(Y_{2,2}(\vec r)-Y_{2,-2}(\vec r)\Big) \;.
	\end{equation}

The Schrodinger
equation in the space spanned by the given set of Gaussians is represented
by the generalized matrix eigenvalue problem,
	\begin{equation}
H^{(l)}c^{(l)}=E_l N^{(l)}c^{(l)} \;,
	\end{equation}
where $N^{(l)}$ is the overlap matrix,
	\begin{equation}
N_{ij}^{(l)}=\langle G_i^{(l)}|G_j^{(l)}\rangle \,,
	\end{equation}
$H^{(l)}$ is the Hamiltonian matrix,
	\begin{equation}
H_{ij}^{(l)}=\langle G_i^{(l)}|\hat H|G_j^{(l)}\rangle \,,
	\end{equation}
and $c^{(l)}=\{c_1^{(l)},\dots,c_n^{(l)}\}$ is the column of the
expansion coefficients.  The matrix elements are calculated using the
derived formulae.

\begin{figure}
\centerline{
\includegraphics{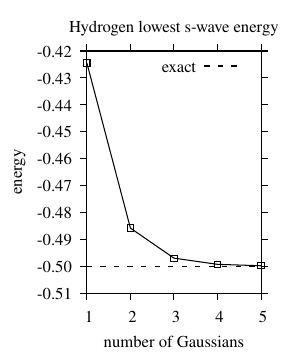}
\includegraphics{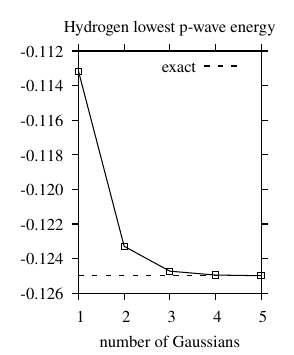}
\includegraphics{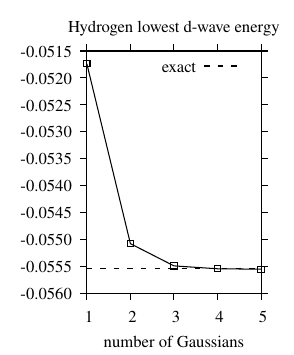}
}\label{fig-H}
\end{figure}

The generalized matrix eigenvalue problem is solved using the standard
method (via Cholesky decomposition of the overlap matrix). The parameters
$\alpha_i$ of the Gaussians are tuned to minimize the lowest eigenvalue by
a gradient-descent method.  The results of the calculations are shown on
figure~(\ref{fig-H}) where the lowest energies for s-, p-, and d-waves
are shown as functions of the number of Guassians in the variational
basis. The exact energies\footnote{the exact energies are, in Hartee
units, $E_{l=0}=-\frac12$, $E_{l=1}=-\frac1{8}$, $E_{l=2}=-\frac1{18}$.}
are reproduced within 4 decimal digits with the basis of 5 Gaussians.

\section{Conclusion}
We have derived analytic matrix elements---overlap, kinetic energy,
and Coulomb potential---for correlated Gaussians with tensor
pre-factors. Tensor pre-factor Gaussians might be useful in certain
nuclear and particle physics applications, in particular when pions are
included explicitly.  We have done a quick test of the derived formulae
by applying them to the p- and d-waves of the hydrogen atom.

\end{document}